\documentclass[11pt]{article}
\usepackage{verbatim}
\usepackage{amsmath}
\usepackage{amsfonts}
\usepackage{amstext}
\usepackage{amssymb}
\usepackage{graphicx}

\ifx\pdfoutput\@undefined\usepackage[usenames,dvips]{xcolor}
\else\usepackage[usenames,dvipsnames]{xcolor}
\IfFileExists{pdfcolmk.sty}{\usepackage{pdfcolmk}}{}
\fi
\usepackage[plainpages=false,pdfpagelabels,pagebackref=false,naturalnames=true,hyperindex=true,pdftitle={Correspondence and Independence of Numerical Evaluations of Algorithmic Information Measures},pdfauthor={Hector Zenil}]{hyperref}
\hypersetup{colorlinks=true,
urlcolor=Cerulean,linkcolor=BrickRed,citecolor=RoyalBlue,a4paper,
 pdfpagemode=None,
 pdfstartview=FitH}
\usepackage[all]{hypcap}

\newenvironment{myenumerate}{
\begin{enumerate}
 \setlength{\itemsep}{1pt}
 \setlength{\parskip}{0pt}
 \setlength{\parsep}{0pt}}{\end{enumerate}
}

\begin{document}

\title{Correspondence and Independence of Numerical Evaluations of Algorithmic Information Measures\thanks{Algorithmic Nature Group URL: \url{http://www.algorithmicnature.org}. Online Algorithmic Complexity Calculator (OACC) URL: \url{http://www.complexitycalculator.com}.}}
\author{Fernando Soler-Toscano$^1$, Hector Zenil$^2$\footnote{Corresponding author: \url{hectorz@labores.eu}}, Jean-Paul Delahaye$^3$\\and Nicolas Gauvrit$^4$\\
\small{$^1$ Grupo de L\'ogica, Lenguaje e Informaci\'on, Universidad de Sevilla, Spain.}\\
\small{$^2$ Department of Computer Science, University of Sheffield, UK.}\\
\small{$^3$ Laboratoire d'Informatique Fondamentale de Lille, France.}\\
\small{$^4$ LDAR, Universit\'e de Paris VII, Paris, France.}}

\date{}
\maketitle

\begin{abstract}
We show that real-value approximations of Kolmogorov-Chaitin ($K_\textit{m}$) using the algorithmic Coding theorem as calculated from the output frequency of a large set of small deterministic Turing machines with up to 5 states (and 2 symbols), is in agreement with the number of instructions used by the Turing machines producing $s$, which is consistent with strict integer-value program-size complexity. Nevertheless, $K_\textit{m}$ proves to be a finer-grained measure and a potential alternative approach to lossless compression algorithms for small entities, where compression fails. We also show that neither $K_\textit{m}$ nor the number of instructions used shows any correlation with Bennett's Logical Depth $LD(s)$ other than what's predicted by the theory. The agreement between theory and numerical calculations shows that despite the undecidability of these theoretical measures, approximations are stable and meaningful, even for small programs and for short strings. We also announce a first Beta version of an Online Algorithmic Complexity Calculator (OACC), based on a combination of theoretical concepts, as a numerical implementation of the \emph{Coding Theorem Method}.\\

\textbf{Keywords:} Coding Theorem Method; Kolmogorov complexity; Solomonoff-Levin algorithmic probability; program-size complexity; Bennett's Logical Depth; small Turing machines.
\end{abstract}

\section{Introduction}

Kolmogorov complexity (also known as Kolmogorov-Chaitin or program-size complexity) is recognized as a fundamental concept, but it is also often thought of as having little or no applicability because it is not possible to provide stable numerical approximations for finite---particularly short---strings by using the traditional approach, namely lossless compression algorithms. We advance a method that can overcome this limitation, and which, though itself limited in ways both theoretical and numerical, nonetheless offers a means of providing sensible values for the complexity of short strings, complementing the traditional lossless compression method that works well for long strings. This is done at the cost of massive numerical calculations and through the application of the Coding theorem from algorithmic probability theory that relates the frequency of production of a string to its Kolmogorov complexity.

Bennett's logical depth, on the other hand, is a measure of the complexity of strings that, unlike Kolmogorov complexity, measures the \emph{organized} information content of a string. In~\cite{zenilld} an application inspired by the notion of logical depth was reported in the context of the problem of image classification. However, the results in this paper represent the first attempt to provide direct numerical approximations of logical depth.

The independence of the two measures-- Kolmogorov complexity and logical depth-- which has been established theoretically, is also numerically tested and confirmed in this paper. Our work is in agreement with what the theory predicts, even for short strings--despite the limitations of our approach. Our attempt to apply these concepts to practical problems (detailed in a series of articles (see e.g.~\cite{delahayezenil,zenilld})) is novel, and they are indeed proving to have interesting applications where evaluations of the complexity of finite short strings are needed~\cite{kolmo2d}.

In Sections~\ref{kolmochap},~\ref{ld},~\ref{codingchap} and~\ref{formal}, we introduce the measures, tools and formalism used for the method described in Section~\ref{Dist}. In Section~\ref{comparison}, we report the numerical results of the evaluation and analysis of the comparisons among the various measures, particularly the connection between number of instructions, integer valued program-size complexity, Kolmogorov complexity approximated by means of the Coding Theorem method, and logical depth.

\section{Kolmogorov-Chaitin complexity}
\label{kolmochap}

When researchers have chosen to apply the theory of algorithmic information (AIT), it has proven to be of great value despite initial reservations~\cite{chaitinestonia}. It has been successfully applied, for example, to DNA false positive repeat sequence detection in genetic sequence analysis~\cite{rivals}, in distance measures and classification methods~\cite{cilibrasi}, and in numerous other applications~\cite{li}. This effort has, however, been hamstrung by the limitations of compression algorithms--currently the only method used to approximate the Kolmogorov complexity of a string--given that this measure is not computable. 

Central to AIT is the basic definition of plain algorithmic (Kolmogorov-Chaitin or program-size) complexity~\cite{kolmo,chaitin}:

\begin{equation}
\label{kolmo}
K_U(s) = \min \{|p|, U(p)=s\}
\end{equation}

Where $U$ is a universal Turing machine and $p$ the program that, running on $U$, produces $s$. Traditionally, the way to approach the algorithmic complexity of a string has been by using lossless compression algorithms. The result of a lossless compression algorithm applied to $s$ is an upper bound of the Kolmogorov complexity of $s$. Short strings, however, are difficult to compress in practice, and the theory does not provide a satisfactory solution to the problem of the instability of the measure for short strings.

The invariance theorem, however, guarantees that complexity values will only diverge by a constant $c$ (e.g. the length of a compiler or a translation program).\\

\noindent \textbf{Invariance Theorem} (\cite{calude,li}): If $U_1$ and $U_2$ are two universal Turing machines, and $K_{U_1}(s)$ and $K_{U_2}(s)$ the algorithmic complexity of $s$ for $U_1$ and $U_2$ respectively, there exists a constant $c$ such that:  

\begin{equation}
\label{invariance}
| K_{U_1}(s) - K_{U_2}(s) | < c
\end{equation} 

Hence the longer the string, the less important $c$ is (i.e. the choice of programming language or universal Turing machine). However, in practice $c$ can be arbitrarily large, thus having a very great impact on the stability of Kolmogorov complexity approximations for short strings.

\section{Bennett's Logical Depth}
\label{ld}

A measure of the structural complexity (i.e. richness of structure and organization) of a string can be arrived at by combining the notions of algorithmic information and time complexity. According to the concept of logical depth \cite{bennett,bennett2}, the complexity of a string is best defined by the time that an unfolding process takes to reproduce the string from its shortest description. While Kolmogorov complexity is related to compression length, Bennett's logical depth is related to decompression time.

A typical example that illustrates the concept of logical depth, underscoring its potential as a measure of complexity, is a sequence of fair coin tosses. Such a sequence would have a high information content (Kolmogorov complexity) because the outcomes are random, but it would have no structure because it is easily generated. The string 1111\ldots1111 would be equally shallow as measured by logical depth. Its compressed version, while very small, requires little time to decompress into the original string. In contrast, the binary expansion of the mathematical constant $\pi$ is not shallow, because though highly compressible and hence having a low Kolmogorov complexity, it requires non-negligible computational time to produce arbitrary numbers of digits from its shortest program (or indeed from any short program computing the digits of $\pi$). A detailed explanation pointing out the convenience of the concept of logical depth as a measure of organized complexity as compared to plain algorithmic complexity, which is what is usually used, is provided in \cite{delahaye}. For finite strings, one of Bennett's formal approaches to the logical depth of a string is defined as follows:\\

Let $s$ be a string and $d$ a significance parameter. A string's depth at significance $d$ is given by
\begin{equation}
\label{ld}
LD_d(s)=\min\{t(p) : (|p|-|p^*| < d) \textit{ and } (U(p) = s)\}
\end{equation}

\noindent with $|p^*|$ the length of the shortest program for $s$, (therefore $K(s)$). In other words, $LD_d(s)$ is the least time $t$ required to compute $s$ from a $d$-incompressible program $p$ on a universal Turing machine $U$. 

Each of the three linked definitions of logical depth provided in~\cite{bennett} comes closer to a definition in which near-shortest programs are taken into consideration. In this experimental approach we make no such distinction among significance parameters, so we will denote the logical depth of a string $s$ simply by $LD(s)$.

Like $K(s)$, $LD(s)$ as a function of $s$ is uncomputable. A novel feature of this research is that we provide exact numerical approximations for both measures-- $K(s)$ and $LD(s)$-- for specific short $s$, allowing a direct comparison. This was achieved by running a large set of random Turing machines and finding the smallest and fastest machines generating each output string. Hence these approximations are deeply related to another important measure of algorithmic information theory.


\section{Solomonoff-Levin Algorithmic Probability}
\label{codingchap}

The algorithmic probability (also known as Levin's semi-measure) of a string $s$, is a measure that describes the expected probability of a random program $p$ running on a universal (prefix-free~\footnote{The group of valid programs forms a prefix-free set, that is, no element is a prefix of any other, a property necessary to keep $0 < m(s) < 1$. For details see~\cite{calude}).}) Turing machine $U$ producing $s$. Formally~\cite{solomonoff,levin,chaitin}, 

\begin{equation}
\label{ap}
m(s) = \Sigma_{p:U(p) = s} 1/2^{|p|}
\end{equation}
i.e. the sum over all the programs for which $U$ with $p$ outputs $s$ and halts. 

Levin's semi-measure $m(s)$ defines a distribution known as the \emph{Universal Distribution}~\cite{kircher}. It is important to notice that the value of $m(s)$ is dominated by the length of the smallest program $p$ (when the denominator of $m$ reaches its largest value). The length of the smallest program $p$ that produces the string $s$ is $K(s)$. The semi-measure $m(s)$ is therefore also uncomputable, because for every $s$, $m(s)$ requires the calculation of $2^{-K(s)}$, involving $K$, which is itself uncomputable. An extension of $m(s)$ to non-binary alphabets is natural. More formally, $m_2(s)$ can be associated with the original definition for binary strings. However, one may want to extend $m_2(s)$ to $m_k(s)$, in which case for every $k$, the function $s \rightarrow m_k(s)$ is semi-computable (for the same reason that $s \rightarrow m_2(s)$ is uncomputable).

An alternative~\cite{delahayezenil} to the traditional use of compression algorithms to approximate $K$ can be derived from a fundamental theorem that establishes the exact connection between $m(s)$ and $K(s)$.\\

\noindent \textbf{Coding Theorem} (Levin~\cite{levin}):
\begin{equation}
\label{precoding}
|-\log_2 m(s) - K(s)| < c
\end{equation}

This theorem posits that if a string has many long descriptions it also has a short one. It elegantly connects frequency to complexity, more specifically the frequency (or probability) of occurrence of a string with its algorithmic (Kolmogorov) complexity. The Coding theorem implies that~\cite{cover,calude} one can calculate the Kolmogorov complexity of a string from its frequency~\cite{zenil2007,delahaye2007,thesis,delahayezenil}, simply rewriting the formula as:

\begin{equation}
\label{coding}
K_\textit{m}(s)=-\log_2 m(s) + O(1)
\end{equation}

Where we will use $K_\textit{m}$ to indicate that $K$ has been approximated by means of $m$ through the Coding theorem. An important property of $m$ as a semi-measure is that it dominates any other effective semi-measure $\mu$, because there is a constant $c_\mu$ such that for all $s$, $m(s) \geq c_\mu\mu(s)$. For this reason $m(s)$ is often called a \emph{Universal Distribution}~\cite{kircher}.

\section{Deterministic Turing machines}
\label{formal}

The ability of a universal Turing machine to simulate any algorithmic process\footnote{Under Church's hypothesis.} has motivated and justified the use of universal Turing machines as the language framework within which definitions and properties of mathematical objects are given and studied.

However, it is important to describe the formalism of a Turing machine, because exact values of algorithmic probability for short strings will be approximated under this model, both for $K(s)$-- through $m(s)$ (denoted by $K_\textit{m}$), and for $K(s)$--in terms of the number of instructions used by the smallest Turing machine producing $s$.\\

Consider a Turing machine $T$ with alphabet $\Sigma=\{0,1\}$ symbols, $\{1,2, \ldots, n\}$ states and an additional halting state denoted by $0$ (as defined by Rado in his original Busy Beaver paper~\cite{rado}). At the outset the Turing machine is in its initial state $1$.

The machine runs on a $2$-way unbounded tape. Its behavior is determined by the transition function $\delta_T$. So, at each step: 
\begin{myenumerate}
\item the machine's current ``state'' (instruction) $s$; and
\item the tape symbol the machine's head is scanning $k$
\end{myenumerate}
define the transition $\delta_T(s,k) = (s',k',d)$ with 
\begin{myenumerate}
\item a unique symbol $k'$ to write (the machine can overwrite a $1$ on a $0$, a $0$ on a $1$, a $1$ on a $1$, and a $0$ on a $0$);
\item a direction $d$ to move in: $-1$ (left), $1$ (right) or $0$ (none, when halting); and
\item a state $s'$ to transition into (which may be the same as the one it was in).
\end{myenumerate}

The machine halts if and when it reaches the special halt state $0$. There are $(4n + 2)^{2n}$ Turing machines with $n$ states and $2$ symbols according to the formalism described above, as there are $2n$ entries in the transition table and any of them may have $4n+2$ possible instructions: there are $2$ halting instructions (writing `0' and `1') and $4n$ non-halting instructions ($2$ movements, $2$ possible symbols to write and 
$n$ states). The output string is taken from the number of contiguous cells on the tape the head of the halting $n$-state machine has gone through. A Turing machine is considered to produce an output string only if it halts. The output is what the machine has written on the tape.

\section{The \emph{Coding Theorem Method}}
\label{Dist}

In order to arrive at an approximation of $m(s)$, a method based on the Coding theorem was advanced in~\cite{delahayezenil}. It is captured in the following function. Let $T$ be a Turing machine in $(n, m)$ with empty input. Then:

\begin{equation}
\label{D}
D(n, m)(s)=\frac{|\{T\in(n, m) : T \textit{ produces } s\}|}{|\{T \in(n, m) : T \textit{ halts }\}|}
\end{equation}

Where $|E|$ denotes the number of elements of $E$. Let $m$ be fixed. It has been proved~\cite{thesis,delahayezenil} that the function $n \rightarrow D(n, m)$ is non-computable (due to the denominator). However, $D(n, m)$ for fixed and small values $n$ and $m$ is computable for values of the Busy Beaver problem~\cite{rado} that are known. For $n=4$, for example, the Busy Beaver function $S$ tells us that $S(4, 2)=107$ ~\cite{brady}, so given a Turing machine with 4 states running on a blank tape that hasn't halted after 107 steps, we know it will never stop. 

More generally, for every string $s$ (with alphabet $\{0, 1, \ldots, m-1\}$) one can compute a sequence $D(n, m) (s, t)$ which converges to $D(n, m)(s)$ when $t \rightarrow \infty$. For $D(n, m)(s, t)$ we compute for $t$ steps all $m$-Turing machines with $n$ states (there is a finite number of them) and compute the quotient for $D(n, m)(s)$ for machines that halted before $t$ steps. Since $D(n, m)(s, t)$ converges for every $s$ to $D(n ,m)(s)$ ($m$ fixed, $n$ fixed, $s$ fixed), the value of $D(n,m)(t)$ converges for fixed $m$ and $n$. In this specific sense $D(n,m)$ is approachable, even if $D(n, m)(s, t)$ as a function of increasing time $t$ may increase when a machine produces $s$, or decrease when a machine halts without producing $s$. By the invariance theorem (Eq.~\ref{invariance}) and the Coding theorem (Eq.~\ref{coding}), $D(n ,m)(s)$ is guaranteed to converge to $m(s)$.

Exact values of $D(n, m)(s, t)$ were previously calculated~\cite{thesis,delahayezenil} for $m=2$ symbols and $n=1, \ldots, 4$ states for which the Busy Beaver values are known. That is, a total of 36, 10\,000, 7\,529\,536 and 11\,019\,960\,576 Turing machines respectively. The distributions were very stable and are proving to be capable of delivering applications that are also in agreement with results from lossless compression algorithms~\cite{kolmo2d} for boundary cases (where both methods can be applied), hence validating the utility of both methods (compression being largely validated by its plethora of applications and by the fact that it achieves an approximation of $K$, see e.g.~\cite{li}). The chief advantage of the Coding Theorem method, however, is that it is capable of dealing with short entities (unlike compression algorithms, which are designed for large entities).

There are 26\,559\,922\,791\,424 Turing machines with 5 states and 2 symbols, and the values of Busy Beaver functions for these machines are unknown. In what follows we describe how we proceeded. Calculating $D(5, 2)$ is an improvement on our previous numerical evaluations and provides a larger data set to work with, allowing us to draw more significant statistical conclusions vis-\`a-vis the relation between these calculations and strict integer value program-size complexity, as well as to make a direct comparison to Bennett's logical depth.

\subsection{Reduction techniques}
\label{sec:some-reductions}

We did not run all the Turing machines with 5 states to produce $D(5,2)$,
because one can take advantage of symmetries and anticipate some of the
behavior of the Turing machines directly from their transition tables
without actually running them (this is impossible generally due to
the halting problem, but some reductions are possible). If $(n,m)$ is
the set of Turing machines with $n$ states and $m$ symbols, as defined
above, we reduce it to: 
\begin{equation*}
  \label{eq:1}
  Red(n,m) = \{T\in(n,m) \mid \delta_T(1,0) = (s,k,1), 
  s\notin\{0,1\},  k\in\{0,\cdots,m-1\}
\}
\end{equation*}
where $\delta_T$ is the transition function of $T$. So
$Red(n,m)$ is a subset of $(n,m)$, with machines with the transition 
corresponding to initial state $1$ and symbol $0$ (this is the initial
transition in a `0'-filled blank tape) moving to the right and
changing to a state different from the initial and halting ones. 

For machines with two symbols, 
\begin{equation*}
  |Red(n,2)| = 2(n-1)(4n+2)^{2n-1}
\end{equation*}
as there are $2(n-1)$ different initial transitions (the machine can
write `0' or `1' and move to one of the $n-1$ states in
$\{2,\dots,n\}$), and for the other $2n-1$ transitions there are $4n+2$
possibilities, as in $(n,m)$.

After running $Red(n,2)$ on a `0'-filled tape, the procedure for completing
the output strings so that they reach the frequency they have in $(n,m)$ is:
\begin{itemize}
\item For every $T \in Red(n,2)$, 
  \begin{itemize}
  \item If $T$ halts and produces the output string $s$, add one
    occurrence of $rev(s)$, the reverse of $s$.
  \item If $T$ does not halt, count another non-halting machine. 
  \end{itemize}
  These two completions add the output (or number of non-halting
  machines) of $2(n-1)(4n+2)^{2n-1}$ new machines, one for each
  machine in $Red(n,2)$. These new machines are left-right symmetric
  to the machines in $Red(n,2)$. Formally, this is the set
  \begin{eqnarray*}
    \{T' \mid T\in Red(n,2),\  \delta_{T'}(s,k) & = &
    (s',k',-d)\ \text{iff} \\
    \delta_{T}(s,k) & = &
    (s',k',d),\ \text{for all}\ s\  \text{and}\ k)\}
  \end{eqnarray*}
  When the original machine halts, its symmetric counterpart halts too and produces the
  reversed strings, and if the original machine does not halt, neither
  does the symmetric machine. This way we consider the output of all machines
  with the initial transition moving to the left and to a state not in 
  $\{0,1\}$. 
\item Include $(4n+2)^{2n-1}$ occurrences of string ``1''. This
  corresponds to the machines writing `1' and halting at the initial
  transition. There is just one possible initial transition for these
  machines (move to the halting state, write `1' and remain in the
  initial cell). The other $2n-1$ transitions can have any of the
  $4n+2$ possible instructions.
\item Include $(4n+2)^{2n-1}$ occurrences of string ``0''. This is
  justified as above, for machines writing `0' and halting at the
  initial transition. 
\item Include $4(4n+2)^{2n-1}$ additional non-halting machines,
  corresponding to machines remaining in the initial state in the
  initial transition (these machines never halt, as they remain forever in
  the initial state). There are $4$ initial transitions of this kind,
  as the machine can write $2$ different symbols and move in $2$
  possible directions. 
\end{itemize}
If we sum $|Red(n,2)|$ and the machines considered above, having
completed them in the manner described, we get the output corresponding to
the $(4n+2)^{2n}$ machines in $(n,2)$. 

Moreover, we need the output of those machines starting with a
`0'-filled tape and with a `1'-filled tape. But we do not run any
machine twice, as for every machine $T\in (n,2)$ producing the
binary string $s$ starting with a `1'-filled tape, there is also a 0-1
symmetric machine $T'\in (n,2)$ (where the role of 1 (of 0) in the
transition table of $T'$ is the role of 0 (of 1) in the transition
table of $T$) that when starting with a `0'-filled tape produces
the complement to one of $s$, that is, the result of replacing 
all 0s in s with 1s and all 1s with 0s. So we add the complement to
every one of the strings found and count the non-halting machines twice to obtain
the output of all machines in $(n,2)$ starting both with a `0'-filled
tape and with a `1'-filled tape. 

To construct $D(5,2)$, we ran the $9\,658\,153\,742$ machines in
$Red(5,2)$, which is $4/11$ of $|(5,2)|$. The output strings found in
$Red(5,2)$, together with their frequencies, were completed prior to
constructing $D(5,2)$, following the procedure explained above.

\subsection{Detecting non-halting machines}
\label{sec:detect-non-halt}

It is useful to avoid running machines that we can easily determine
will not stop. These machines will consume the runtime without
yielding an output. As we have shown above, we can avoid generating
many non-halting machines. In other cases, we can detect them at
runtime, by setting appropriate filters. The theoretical limit of the
filters is the halting problem, which means that they cannot be
exhaustive. But a practical limit is imposed by the difficulty of
checking some filters, which takes up more time than the runtime that
is saved.

We have employed some filters that have proven useful. Briefly,
these are:

\begin{itemize}
\item \textbf{Machines without transitions to the halting
    state}. While the transition table is being filled, the simulator
  checks to ascertain whether there is some transition to the halting
  state. If not, it avoids running it.
\item \textbf{Escapees}. These are machines that at some stage begin
  running forever in the same direction. As they are always reading
  new blank symbols, as soon as the number $c$ of non-previously visited
  positions is greater than the number $n$ of states, we know that they
  will not stop, because the machines have necessarily entered an
  infinite loop. Given that $c>n$, while visiting the last $c$ new cells,
  some of the $n$ states have been repeated, and will repeat
  forever, as the machine's behavior is deterministic. 
\item \textbf{Cycles of period two}. These cycles are easy to
  detect. They are produced when in steps $t$ and $t+2$ the tape is
  identical and the machine is in the same state and the same
  position. When this is the case, the cycle will be repeated
  infinitely.
\end{itemize}

These filters were implemented in our C++ simulator, which also uses
the reduced enumeration of Section~\ref{sec:some-reductions}. To test
them we calculated $D(4,2)$ with the simulator and compared the output to
the list that was computed in~\cite{delahayezenil}, arriving at
exactly the same results, and thereby validating our reduction
techniques.

Running $(4,2)$ without reducing the enumeration or detecting
non-halting machines took 952 minutes. Running the reduced enumeration
with non-halting detectors took 226 minutes.

\subsection{Setting the runtime}
\label{sec:setting-runtime}

The Busy Beaver for Turing machines with 4 states is known to be 107 steps  \cite{brady}, that is, any Turing machine with 2 symbols and 4 states running longer than 107 steps will never halt. However, the exact number is not known for Turing machines with 2 symbols and 5 states, although it is believed to be 47\,176\,870, as there is a candidate machine that runs for this length of time and halts and no machine with a greater runtime has yet been found. 

So we decided to let the machines with 5 states run for 4.6 times the 
Busy Beaver value for 4-state Turing machines (for 107 steps), knowing 
that this would constitute a sample significant enough to capture the 
behavior of Turing machines with 5 states. The chosen runtime was 
rounded to 500 steps, which was used to construct the output frequency
 distribution for $D(5,2)$. 

Not all 5-state Turing machines have been used to build $D(5,2)$, since only the output of machines that halted at or before 500 steps was taken into consideration.
 As an experiment to ascertain how many machines we were leaving out, we ran 
$1.23 \times 10^{10}$ random Turing machines for up to 5000 steps. Among
 these, only 50 machines halted after 500 steps and before 5000 (that 
is, a fraction less than $1.75164\times 10^{-8}$, because in the reduced 
enumeration we don't include those machines that halt in one step or 
that we know won't halt before we generate them, so it's a 
smaller fraction), with the remaining 1\,496\,491\,379 machines not halting 
at 5000 steps. As far as these are concerned--and given that the Busy Beaver values for 5 states are unknown--we do not know after how many steps they would
 eventually halt, if they ever do. According to the following analysis, our
 election of a runtime of 500 steps therefore provides a good estimation of
$D(5,2)$.


The frequency of runtimes of (halting) Turing machines has theoretically been proven to drop exponentially  \cite{calude2}, and our experiments are closer to the theoretically predicted behavior. To estimate the fraction of halting machines that were missed because Turing machines with 5 states were stopped after 500 steps, we hypothesize that the number of steps $S$ a random halting machine needs before halting is an exponential random variable, defined by $\forall k\geq 1, P(S=k)\propto e^{-\lambda k}.$ We do not have direct access to an evaluation of $P(S=k)$, since we only have data for those machines for which $S\leq 5000$. But we may compute an approximation of $P(S=k | S\leq 5000)$, $1\leq k \leq 5000$, which is proportional to the desired distribution.

 A non-linear regression using ordinary least-squares gives the approximation $P(S=k | S\leq 5000)=\alpha e^{-\lambda k}$ with $\alpha = 1.12$ and $\lambda = 0.793$. The residual sum-of-squares is $3.392\times 10^{-3}$; the number of iterations with starting values $\alpha =0.4$ and $\lambda =0.25$ is nine. The model's $\lambda$ is the same $\lambda$ appearing in the general law $P(S=k)$, and may be used to estimate the number of machines we lose by using a 500 step cut-off point for running time: $P(k> 500)\approx e^{-500\lambda}\approx 6\times 10^{-173}$. This estimate is far below the point where it could seriously impair our results: the less probable (non-impossible) string according to $D(5,2)$ has an observed probability of $1.13\times 10^{-9}$.

 Although this is only an estimate, it suggests that missed machines are few enough to be considered negligible.

\section{Comparison of $K_\textit{m}$ with the number of instructions used and Logical Depth}
\label{comparison}

We now study the relation of $K_\textit{m}$ to the minimal number of instructions used by a Turing machine producing a given string, and to Bennett's concept of logical depth. As expected, $K_\textit{m}$
 shows a correlation with the number of instructions used but not with logical depth.

\subsection{Relating $K_\textit{m}$ to the number of instructions used}
\label{sec:numb-used-instr}

First, we are interested in the relation of $K_\textit{m}(s)$ to the minimal 
number of instructions that a Turing machine producing a string $s$ 
uses. Machines in $D(5,2)$ have a transition table with 10 entries,
 corresponding to the different pairs $(n, m)$, with $s$ one of the five 
states and $m$ either ``0'' or ``1''. These are the 10 instructions that 
the machine can use. But for a fixed input not all instructions are 
necessarily used. Then, for a blank tape, not all machines that halt use 
the same number of instructions. The simplest cases are machines 
halting in just one step, that is, machines whose transition for 
$(init\_state, blank\_symbol)$ goes to the halting state, producing a 
string ``0'' or ``1''. So the simplest strings produced in $D(5, 2)$ are 
computed by machines using just one instruction. We expected a
 correlation between the $K_\textit{m}$-complexity of the strings and the number of instructions used. As we show, the following experiment confirmed this. 

\begin{figure}[htbp!]
  \centering
  \includegraphics[width=10cm]{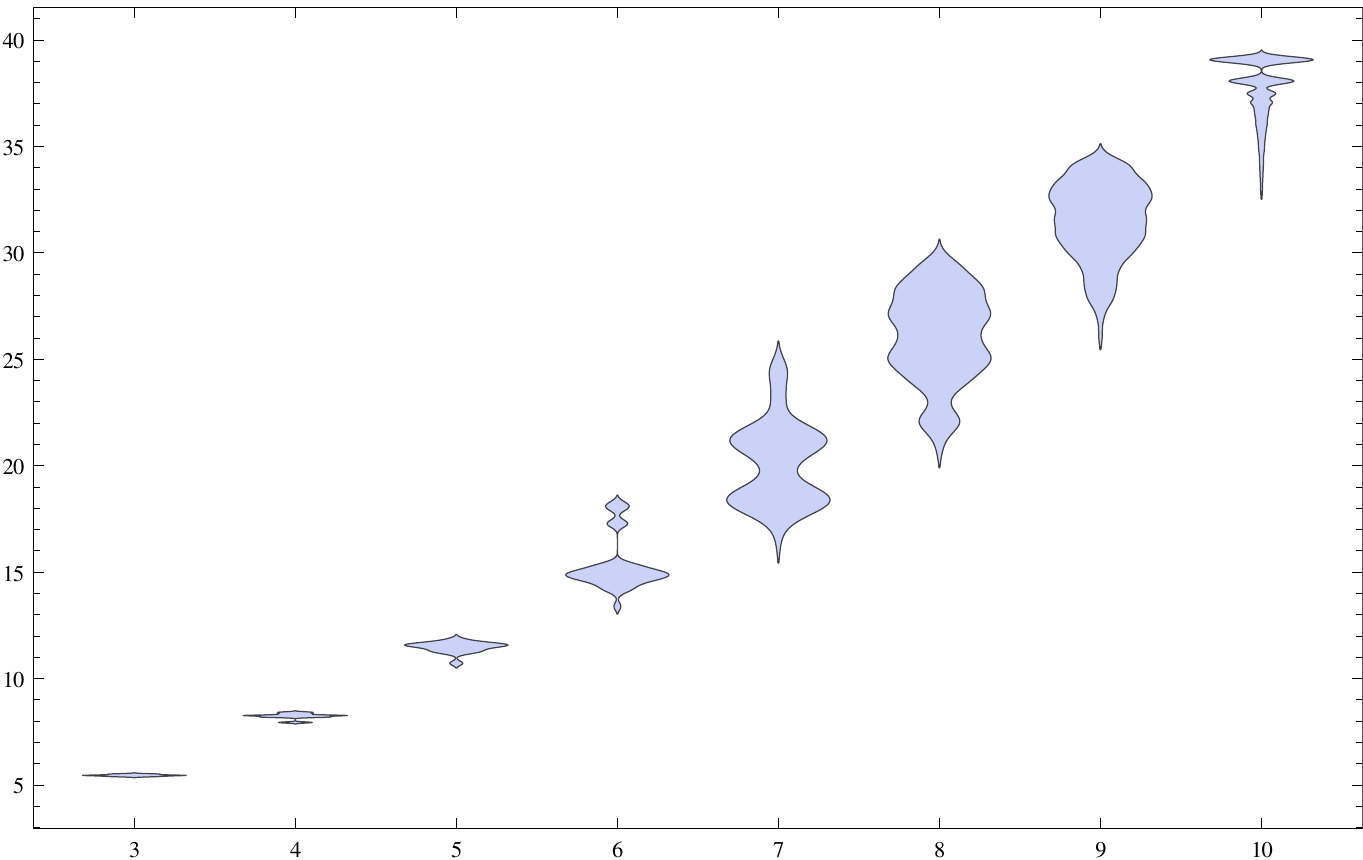}
  \caption{Distribution chart of $K_m$ values according to the minimum number of instructions required. Each ``drop-like" distribution is the set of strings that are minimally produced with the same number of instructions (horizontal axis). The more instructions needed to produce the strings, the more complex they are (vertical axis in $K_m$ units).}
  \label{fig:distribIns}
\end{figure}

We used a sample of $2\,836\times 10^9$ random machines in the reduced enumeration for $D(5, 2)$, that is, $29\%$ the total number of 
machines. The output of the sample returns the strings produced by halting machines together with the number of instructions used, the 
runtime and the instructions for the Turing machine (see Fig.~\ref{fig:distribIns}). In order to save space, we only saved the smallest number of instructions found for each string produced, and the smallest runtime corresponding to that particular number of instructions. 

After doing the appropriate symmetry completions we have 99\,584 
different strings, which is to say almost all the 99\,608 strings found in
 $D(5, 2)$. The number of instructions used goes from 1 to 10. When 1 
instruction is used only ``0'' and ``1'' are generated, with a $K_\textit{m}$ value of $2.51428$. With 2 instructions, all 2-bit strings are generated, with a $K_\textit{m}$ value of $3.32744$. For 3 or more instructions, 
Fig.~\ref{fig:distribIns} shows the distribution of values of $K_\textit{m}$. Table~\ref{tab:meanKL} shows the mean $K_\textit{m}$ values for the different numbers of instructions used.

\begin{table}[htbp!]
  \centering
  \begin{tabular}{|r|r|r|}\hline
    \textbf{Used inst.} & \textbf{Mean $K_\textit{m}$} & \textbf{Mean Length}\\\hline\hline
    1 & 2.51428   &  1  \\
    2 & 3.32744  &  2  \\
    3 & 5.44828   &  3  \\
    4 & 8.22809  &  4  \\
    5 & 11.4584  &  5  \\
    6 & 15.3018  & 6.17949   \\
    7 & 20.1167 & 7.76515   \\
    8 & 26.0095  & 9.99738    \\
    9 & 31.4463 & 12.6341    \\
    10 & 37.5827  & 17.3038   \\\hline
  \end{tabular}
  \caption{Mean $K_\textit{m}$ and string length for different numbers of instructions used.}
  \label{tab:meanKL}
\end{table}

This accords with our expectations. Machines using a low
 number of instructions can be repeated many times by permuting the 
order of states. So the probability of producing their strings is 
greater, which means low $K_\textit{m}$ values.

\begin{figure}[htbp!]
  \centering
  \includegraphics[width=10cm]{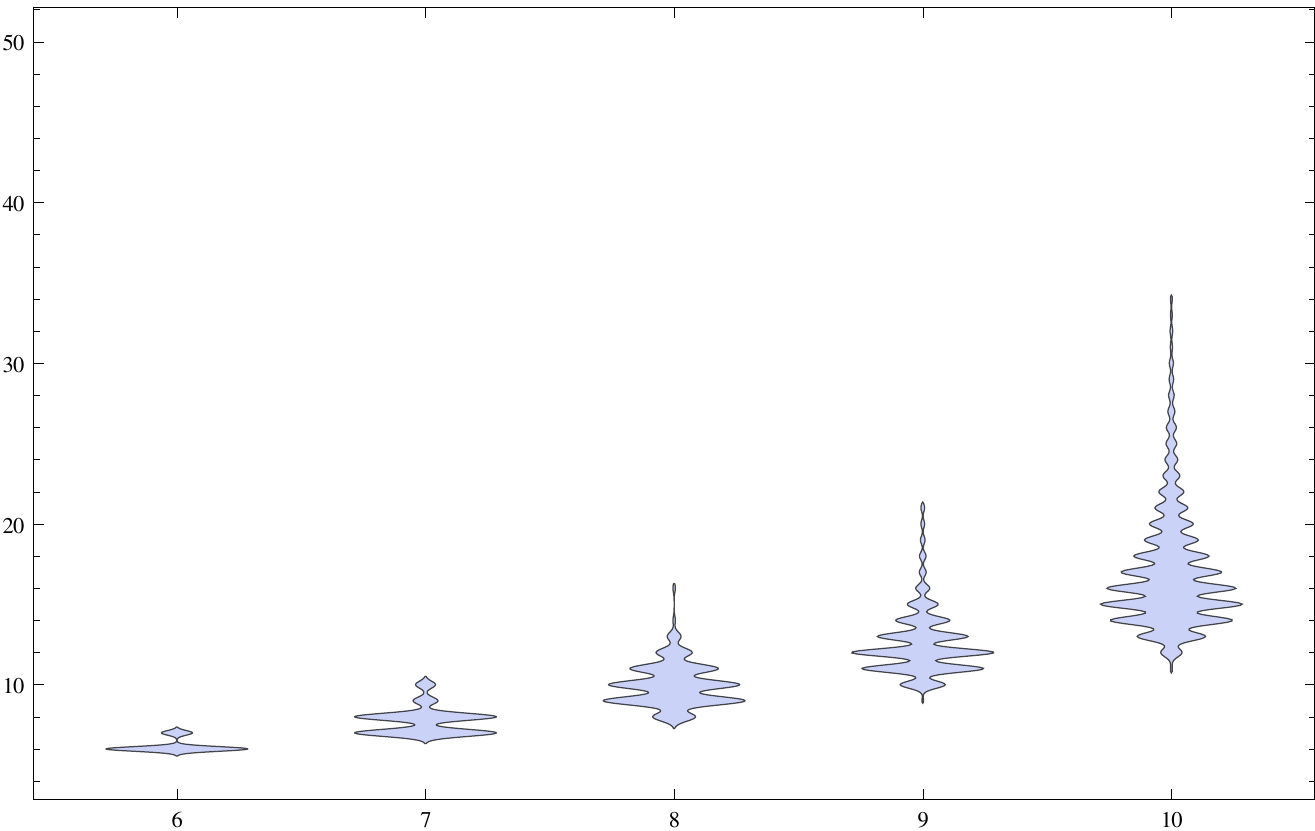}
  \caption{Minimum number of instructions required to produce a string (horizontal axis) and distribution of string lengths (vertical axis).}
  \label{fig:insandL}
\end{figure}

We can also look at the relation between the number of instructions
 used and the length of the strings produced. For $1\leq i\leq 5$, all
 strings of length $i$ are produced by machines using $i$ instructions. For a greater number of instructions used, Fig.~\ref{fig:insandL} shows the distribution of string lengths. Table~\ref{tab:meanKL} shows the mean length for each number of instructions used. 

The correlation $r_{K_{m},N}=0.83$ is a good indicator
for quantifying the apparent relation between $K_{m}$ and the number
$N$ of instructions used, proving a strong positive link. However,
since the length $L$ of outputs is linked with both variables, the
partial correlation $r_{K_{m},N.L}=0.81$ is a better index. This value
indicates a strong relation between $K_{m}$ and $N$, even while
controlling for $L$.

\subsection{Logical Depth and $K_\textit{m}$}
\label{sec:logical-depth-K_textit{m}}

 As explained above, we have also found that the machines which generate each string using the minimum number of instructions also have the minimum runtime. These runtimes are related to Bennett's logical depth ($LD$), as they are the shortest runtimes of the smallest Turing machines producing each
 string in $D(5, 2)$. 
\begin{figure}[htbp!]
  \centering
  \includegraphics[width=10cm]{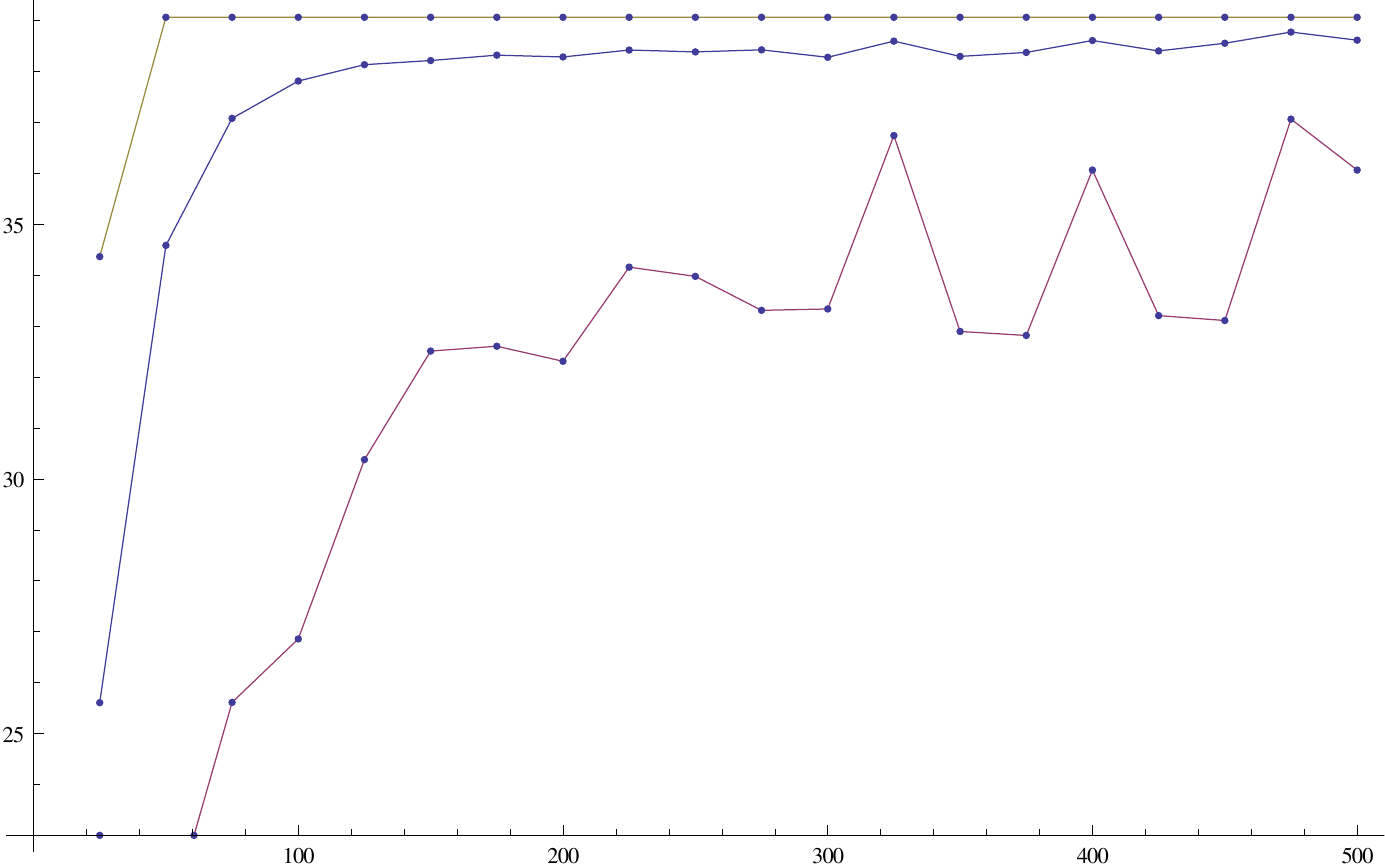}
  \caption{$LD$ and $K_\textit{m}$ (min, mean and max values). The horizontal axis shows the minimum runtime required to produce strings in $D(5,2)$ (divided into intervals of 25 steps) and the vertical axis the minimum, mean and maximum $K_\textit{m}$ values in the interval.}
  \label{fig:LogDepMean}
\end{figure}
\begin{figure}[htbp!]
  \centering
  \includegraphics[width=12cm]{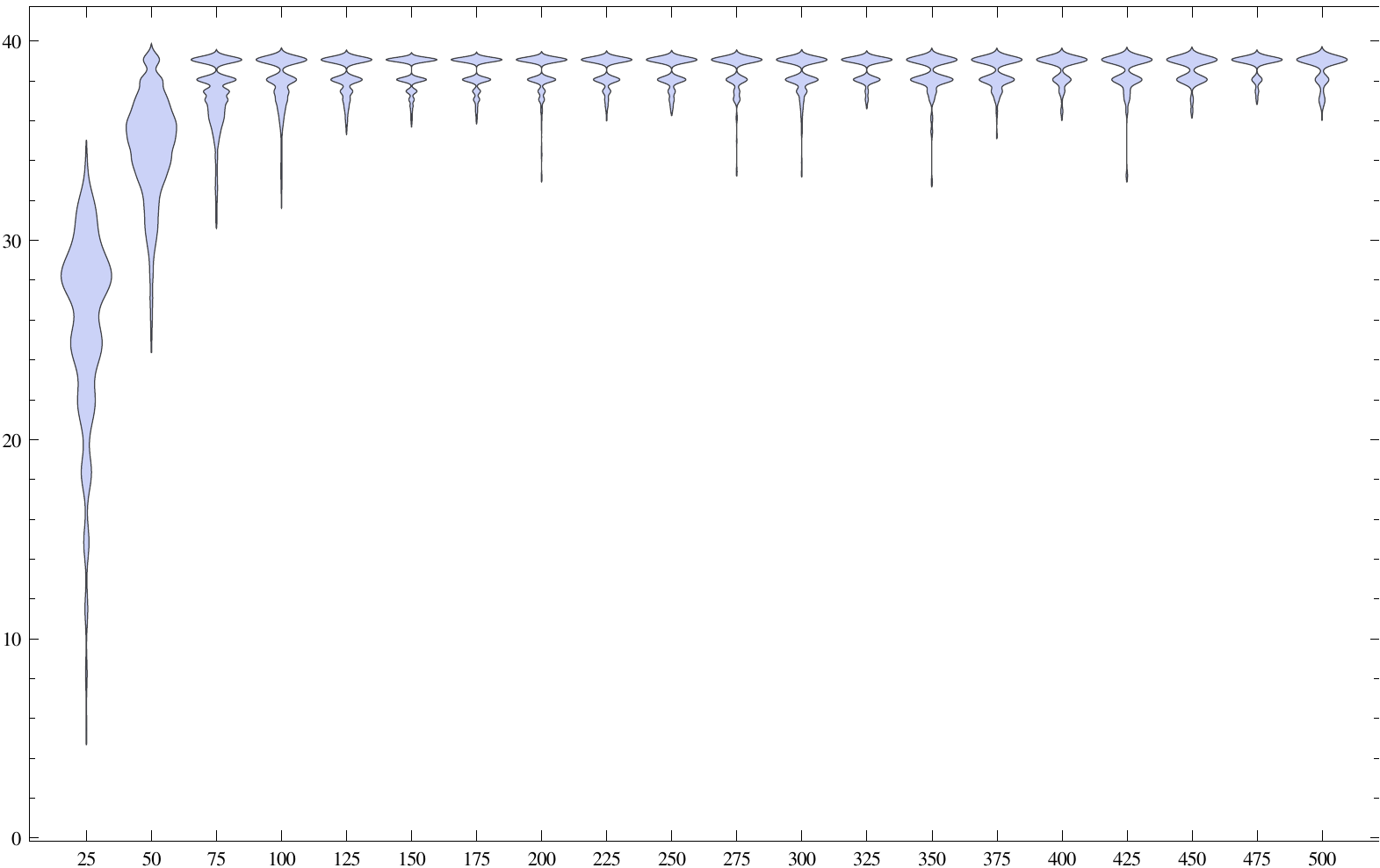}
  \caption{$LD$ and $K_\textit{m}$ (distribution). The vertical axis shows the distribution of $K_\textit{m}$ for each runtime interval (horizontal axis). There is no correlation between $LD$ and $K_\textit{m}$ other than the expected one, viz. that both measures identify simple objects as objects of low complexity.}
  \label{fig:LogDep}
\end{figure}

We have  partitioned the runtime space from 1 to 500 (our 
runtime bound) into 20 groups of equal length 
(25 steps). In order to explore the relation of $K_\textit{m}$ to Bennett's
$LD$ we are interested in the values of $K_\textit{m}$ for the 
strings in each group. Fig.~\ref{fig:LogDepMean} shows the minimum, 
mean and maximum $K_\textit{m}$ values for each of the runtime groups. The same 
information is in Table~\ref{tab:meanKL}. The distribution of $K_\textit{m}$ 
values for the different groups is shown in Fig.~\ref{fig:LogDep}. For each 
interval, the maximum runtime is shown on the horizontal axis. 

\begin{table}[htbp!]
  \centering
  \begin{tabular}{|r|r|r|r|}\hline
    \textbf{Runtime} & \textbf{Min} $K_\textit{m}$ &  \textbf{Mean} $K_\textit{m}$ &
    \textbf{Max} $K_\textit{m}$ \\\hline\hline
1-25 & 2.51428 & 25.6049 & 34.3638\\
26-50 & 21.0749 & 34.5849 & 39.0642\\
51-75 & 25.6104 & 37.0796 & 39.0642\\
76-100 & 26.8569 & 37.8125 & 39.0642\\
101-125 & 30.3777 & 38.1337 & 39.0642\\
126-150 & 32.5096 & 38.2150 & 39.0642\\
151-175 & 32.6048 & 38.3208 & 39.0642\\
176-200 & 32.3093 & 38.2850 & 39.0642\\
201-225 & 34.1573 & 38.4213 & 39.0642\\
226-250 & 33.9767 & 38.3846 & 39.0642\\
251-275 & 33.3093 & 38.4249 & 39.0642\\
276-300 & 33.3363 & 38.2785 & 39.0642\\
301-325 & 36.7423 & 38.5963 & 39.0642\\
326-350 & 32.8943 & 38.2962 & 39.0642\\
351-375 & 32.8163 & 38.3742 & 39.0642\\
376-400 & 36.0642 & 38.6081 & 39.0642\\
401-425 & 33.2062 & 38.4035 & 39.0642\\
426-450 & 33.1100 & 38.5543 & 39.0642\\
451-475 & 37.0642 & 38.7741 & 39.0642\\
476-500 & 36.0642 & 38.6147 & 39.0642\\\hline
  \end{tabular}
  \caption{Extreme and mean $K_\textit{m}$ values for different runtime intervals.}
\end{table}

\begin{figure}[htbp!]
    \centering
    \includegraphics[width=11cm]{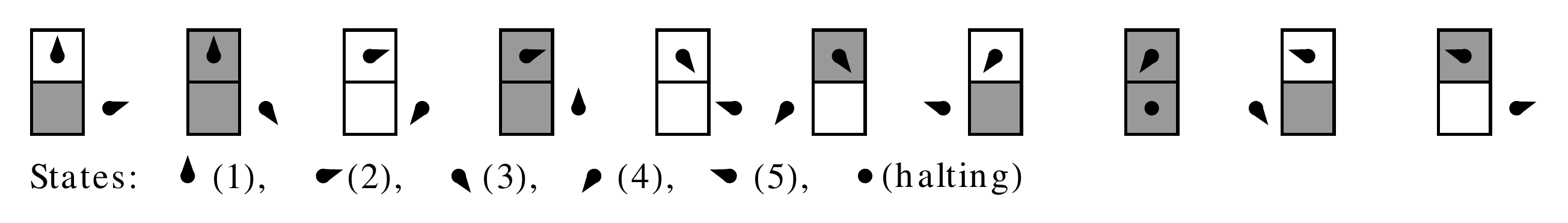}
    \caption{Transitions of a sample machine producing ``0011110001011''.}
    \label{fig:TableLowTime}
  \end{figure}

\begin{figure}[htbp!]
    \centering
    \includegraphics[width=3cm]{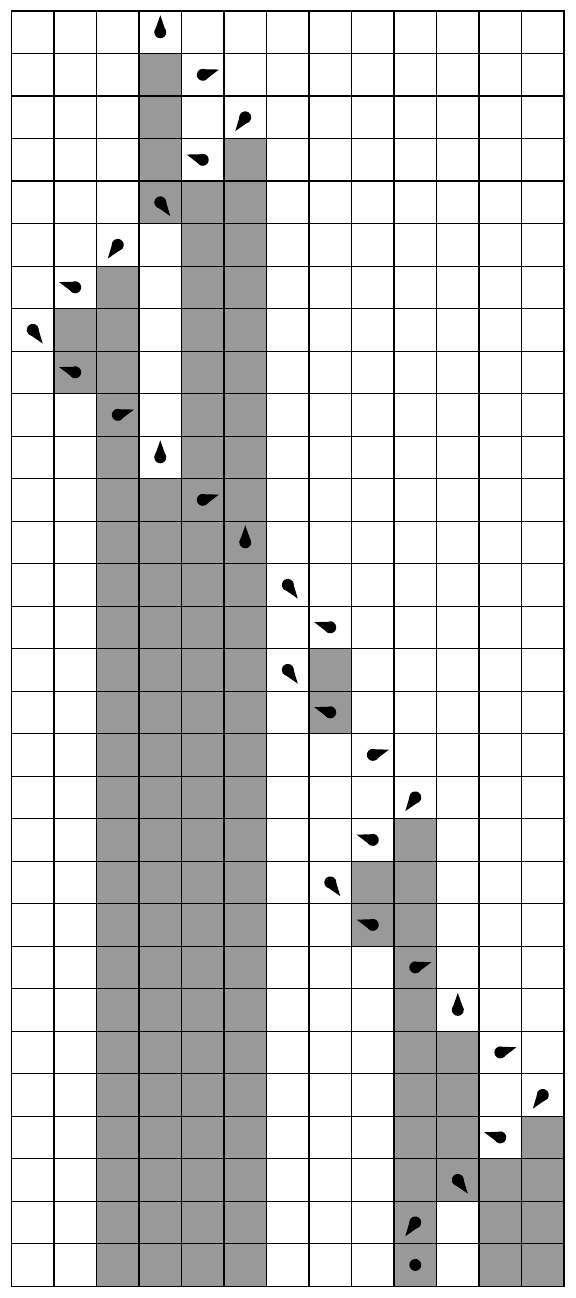}
    \caption{Execution of the machine producing ``0011110001011''. With a low runtime, this sample machine produces a string with high complexity.}
    \label{fig:ExecLowTime}
  \end{figure}

We now provide some examples of the discordance between $K_\textit{m}$ and $LD$. ``0011110001011'' is a string with high $K_\textit{m}$ and low $LD$. Fig.~\ref{fig:TableLowTime} shows the transition table of the smallest machine found producing this string. The runtime is low--just 29 steps (of the 99\,584 different strings found in our sample, only 3\,360 are produced in fewer steps), but it uses 10 instructions and produces a string with complexity $39.0642$. It is the greatest complexity we have calculated for $K_\textit{m}$. Fig.~\ref{fig:ExecLowTime} shows the execution of the machine.  

\begin{figure}[htbp!]
    \centering
    \includegraphics[width=11cm]{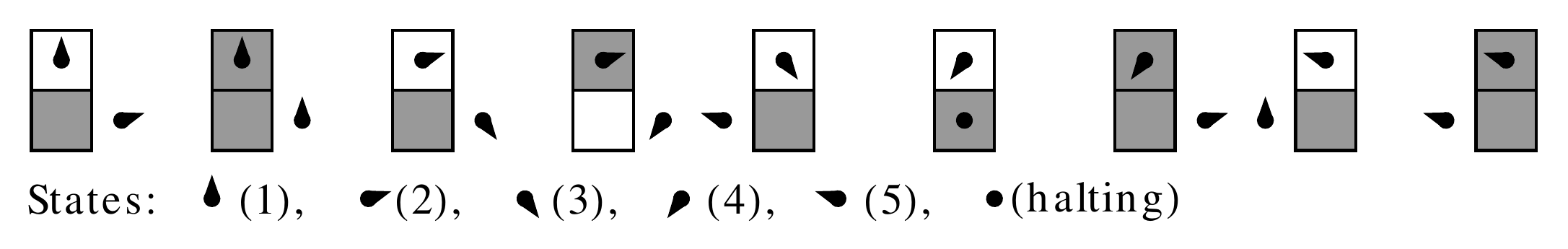}
    \caption{Transition table of a sample machine producing ``$(10)^{20}1$''.}
    \label{TMtableHighTime}
  \end{figure}

\begin{figure}[htbp!]
    \centering
    \includegraphics[width=12cm]{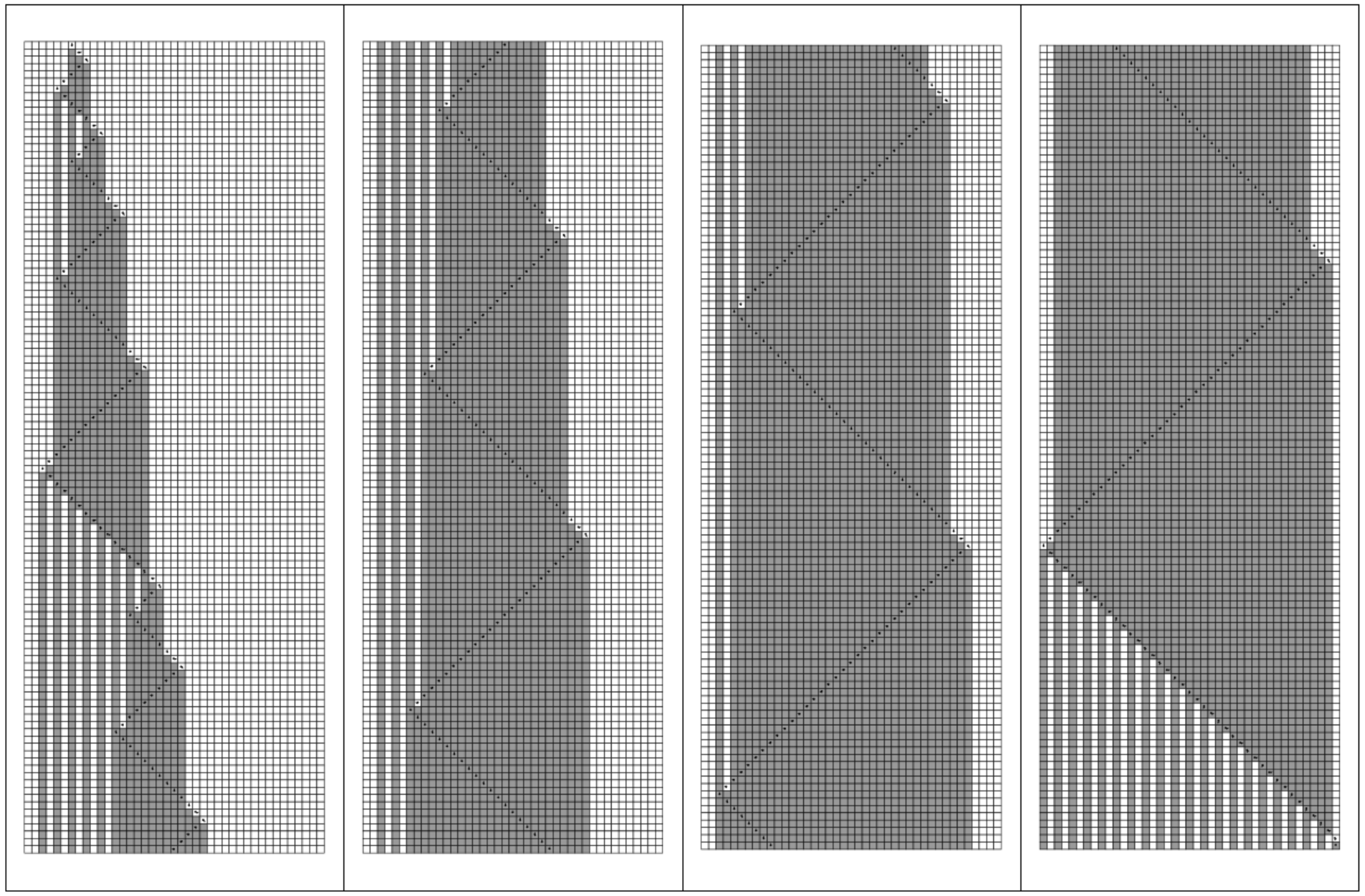}
    \caption{Execution of the sample machine producing ``$(10)^{20}1$''. With a high runtime, it produces a string with low complexity.}
    \label{TMrunHighTime}
  \end{figure}

On the other hand, ``$(10)^{20}1$'' is a string with high $LD$ but a low
$K_\textit{m}$ value. Fig.~\ref{TMtableHighTime} shows the transition  
table of the machine found producing this string, and Fig.~\ref{TMrunHighTime}
 depicts the execution. The machine uses 9 instructions and runs for
 441 steps (only 710 strings out of the 99\,584 strings in our sample
 require more time) but its $K_\textit{m}$ value is $33.11$. This is a low
 complexity if we consider that in $K_\textit{m}$ there are 99\,608 strings
 and that 90\,842 are more complex than this one.  

  
We may rate the overall strength of the relation between $K_{m}$ and
$LD$ by the correlation $r_{K_{m},LD}=0.41$, corresponding to a medium
positive link. As we previously mentioned, however, the fact that the
length $L$ of the strings is linked with both variables may bias our
interpretation. A more relevant measure is thus
$r_{K_{m},LD.L}=-0.06$, a slight negative but with no significant
value between $K_{m}$ and $LD$ once $L$ is controlled.

\section{Concluding remarks}

The results in this paper are important because these measures can be better studied and understood under a specific but widely known general formalism. What we have found is very interesting because it is what one would wish in the best case scenario, stable and reasonable distributions rather than chaotic and unstable ones. The results also suggest that these measures can be applied even if numerically approximated using a specific model of computation.

For example, as we expected, the Kolmogorov-Chaitin complexity evaluated by means of Levin's Coding Theorem from the output distribution of small Turing machines correlates with the number of instructions used but not with logical depth. Logical depth also yields a measure that is different from the measure obtained by considering algorithmic complexity ($K$) alone, and this investigation proves that all these three measures (Kolmogorov-Chaitin Complexity, Solomonoff-Levin Algorithmic Probability and Bennett's Logic Depth) are consistent with theoretical expectations. $K$ as a measure of program-size complexity is traditionally expected to be an integer (the length of a program in bits), but when evaluated through algorithmic probability using the Coding theorem it retrieves non-integer values (still bits). These results confirm the utility of non-integer values in the approximation of the algorithmic complexity of short strings, as they provide finer values with which one can tell apart small differences among short strings--which also means one can avoid the longer calculations that would be necessary in order to tell apart the complexity of very small objects if only integer values were allowed. Thus it also constitutes a complementary and alternative method to compression algorithms. 

An \emph{On-line Algorithmic Complexity Calculator} (or OACC) is now available at
\url{http://www.complexitycalculator.com}. It represents a long-term project to develop an encompassing universal tool implementing some of the measures and techniques described in this paper. It is expected to be expanded in the future as it currently only implements numerical approximations of Kolmogorov complexity and Levin's semi-measure for short binary strings. More measures, more data and better approximations will be gradually incorporated in the future, covering a wider range of objects, such as longer binary strings, non-binary strings and multidimensional arrays (such as images).

\end{document}